\documentclass[english,aps,preprint,nofootinbib]{revtex4}
\usepackage[T1]{fontenc}
\usepackage[latin1]{inputenc}
\usepackage{amsmath}
\usepackage{color}
\usepackage{graphicx}
\usepackage{amssymb}

\makeatletter

\providecommand{\tabularnewline}{\\}

\usepackage{color}

\usepackage{babel}
\makeatother
\begin{document}
\newcommand{\met}{\not\!\! E_{T}}

\newcommand{\pslash}[1]{\not{\!#1}}

\begin{flushright}UCRHEP-T428\\
 MSUHEP-061208\par\end{flushright}

\title{Resummation Effects in the Search of SM Higgs Boson \\
 at Hadron Colliders}

\author{Qing-Hong Cao}

\email{qcao@ucr.edu}

\affiliation{Department of Physics and Astronomy, University of California at
Riverside, Riverside, CA 92521}

\author{Chuan-Ren Chen}

\email{crchen@pa.msu.edu}

\affiliation{Department of Physics and Astronomy, Michigan State University, E.
Lansing, MI 48824}


\begin{abstract}
We examine the soft-gluon resummation effects, including the exact
spin correlations among the final state particles, in the search of
the Standard Model Higgs boson, via the process $gg\to H\to WW/ZZ\to4\,{\rm leptons}$,
at the Tevatron and the LHC. A comparison between the resummation
and the Next-to-Leading order (NLO) calculation is performed after
imposing various kinematics cuts suggested in the literature for the
Higgs boson search. For the $H\to ZZ$ mode, the resummation effects
increase the acceptance of the signal events by about $25\%$, as
compared to the NLO prediction, and dramatically alter various kinematics
distributions of the final state leptons. For the $H\to WW$ mode,
the acceptance rates of the signal events predicted by the resummation
and NLO calculations are almost the same, but some of the predicted
kinematical distributions are quite different. Thus, to precisely
determine the properties of the Higgs boson at hadron colliders, the
soft-gluon resummation effects have to be taken into account. 
\end{abstract}

\pacs{14.80.Bn}

\maketitle

\section{introduction}

Although Standard Model (SM) explains successfully all current high
energy physics experimental data, the mechanism of electoweak spontaneous
symmetry breaking, arising from the Higgs mechanism, has not yet been
tested directly. Therefore, searching for the Higgs boson ($H$) is
one of the most important tasks at the current and future high energy
physics experiments. The negative result of direct search at the LEP2,
via the Higgsstrahlung process $e^{+}e^{-}\to ZH$, poses a lower
bound of $114.1\,{\rm GeV}$ on the SM Higgs boson mass ($M_{H}$)~\cite{Barate:2003sz}.
On the other hand, global fits to electroweak observables prefer $M_{H}\lesssim200\,$GeV
at the $95\%$ confidence level~\cite{lepewwg:2006}, while the triviality
arguments put an upper bound $\sim1\,$TeV~\cite{Hambye:1996wb}.

There is currently an active experimental program at the Tevatron
to directly search for the Higgs boson. The Large Hadron Collider
(LHC) at CERN, scheduled to operate in late 2007, is expected to establish
the existence of Higgs boson if the SM is truly realized in Nature.
At the LHC, the SM Higgs boson is mainly produced through gluon-gluon
fusion process induced by a heavy (top) quark loop. Once being produced,
it will decay into a fermion pair or vector boson pair. The strategy
of searching for the Higgs boson depends on how it decays and how
large the decay branching ratio is. If the Higgs boson is lighter
than $130\,$GeV, it mainly decays into a bottom quark pair ($b\bar{b}$).
Unfortunately, it is very difficult to search for the Higgs boson
in this mode due to the extremely large Quantum Chromodynamics (QCD)
background at the LHC. However, the $H\to\gamma\gamma$ mode can be
used to detect a Higgs boson with the mass below 150\ GeV\ \cite{Froidevaux:1995,Gianotti:1996}
though the decay branching ratio of this mode is quite small, $\sim O(10^{-3})$.
If the Higgs boson mass ($M_{H}$) is in the region of $130\,{\rm GeV}$
to $2M_{Z}$ ($M_{Z}$ being the mass of $Z$ boson), the $H\to ZZ^{*}$
mode is very useful because of its clean collider signature of four
isolated charged leptons. The $H\to WW^{(*)}$ mode is also important
in this mass region because of its large decay branching ratio. When
$M_{H}>2M_{Z}$, the decay mode $H\to ZZ\to\ell^{+}\ell^{-}\ell^{\prime+}\ell^{\prime-}$
is considered as the {}``gold-plated'' mode which is the most reliable
way to detect the Higgs boson up to $M_{H}\sim600\,$GeV because the
backgrounds are known rather precisely and the two on-shell $Z$ bosons
could be reconstructed experimentally. For $M_{H}>600\,{\rm GeV}$,
one can detect the $H\to ZZ\to\ell^{+}\ell^{-}\nu\bar{\nu}$ decay
channel in which the signal appears as a Jacobian peak in the missing
transverse energy spectrum.

The discovery of the Higgs boson relies on how well we understand
the signals and its backgrounds, because one needs to impose optimal
kinematics cuts to suppress the huge backgrounds and enhance the signal
to background ratio ($S/B$). Many works have been done in the literature
to calculate the higher order QCD corrections to the dominant production
process of the Higgs boson $gg\to H$\ \cite{Dawson:1990zj,Djouadi:1991tk,Spira:1995rr,Kramer:1996iq,Harlander:2000mg,Catani:2001ic,Harlander:2001is,Harlander:2002wh,Anastasiou:2002yz,Ravindran:2003um,Catani:2003zt,Moch:2005ky,Ravindran:2006bu}.
\textcolor{black}{In addition to determine the inclusive production
rate of the Higgs boson, an accurate prediction of kinematics of the
Higgs boson is very essential for the Higgs boson search. However,
a fix order calculation cannot reliably predict the transverse momentum
($Q_{T}$) distribution of Higgs boson for the low $Q_{T}$ region
where the bulk events accumulate. This is because of large corrections
of the form ${\rm ln}(Q^{2}/Q_{T}^{2})$ due to non-complete cancellations
of soft and collinear singularities between virtual and real contributions,
where $Q$ is the invariant mass of the Higgs boson. Therefore, one
needs to take into account the effects of the initial state multiple
soft-gluon emissions in order to make a reliable prediction on the
kinematic distributions of the Higgs boson. One approach to achieve
this is to include parton showering~\cite{Sjostrand:1985xi} which
resums the universal leading logs in Monte Carlo event generators,
e.g. HERWIG~\cite{Corcella:2002jc} and PYTHIA~\cite{Sjostrand:2006za},
which are commonly used by experimentalists. The showering process
just depends on the initial state parton and the scale of the hard
process being considered. The advantage is that it could be incorporated
into various physics processes. Recently, an approach to match NLO
matrix element calculation and parton showing Monte Carlo generators,
MC@NLO~\cite{Frixione:2002ik,Frixione:2003ei}, has been proposed.
Another approach is to include correctly the soft-gluon effects is
to calculate an analytical result by using the Collins-Soper-Sterman
(CSS) resummation formalism\ \cite{Collins:1981uk,Collins:1981zc,Collins:1981va,Collins:1984kg}
to resum these large logarithmic corrections to all order in $\alpha_{s}$.
However, in practice the power of logarithms included in Sudakov exponent
depends on which level the fixed order calculation has been performed~\cite{Balazs:2000wv,Berger:2002ut,Kulesza:2003wn,Bozzi:2003jy,Bozzi:2005wk}.
It is very interesting to compare the predictions between parton showering
and resummation calculation and detailed comparisons have been presented
in Ref.~\cite{Balazs:2000wv,Balazs:2000sz,Huston:2004yp,Balazs:2004rd,Dobbs:2004bu}
which concluded that all of the distributions are basically consistent
with each other, except PYTHIA in the small $Q_{T}$ region and HERWIG
in the large $Q_{T}$ region. }

In addition, the spin correlation among the Higgs decay products has
been proved to be crucial to suppress the backgrounds\ \cite{atlas:1999,CMS:2006}.
Hence, an accurate theoretical prediction, which incorporates the
initial state soft-gluon resummation effects and the spin correlations
among the Higgs decay products, is needed. In this paper, we present
such a calculation and study the soft-gluon resummation (RES) effects
on various kinematics distributions of final state particles. Furthermore,
we examine the impact of the RES effects on the acceptance rate of
the signal events with various kinematics cuts (which were suggested
in the literature~\cite{Abazov:2005un,atlas:1999} for Higgs search)
and compare them with the leading order (LO) and NLO predictions\ %
\footnote{The NLO Quantum Electrodynamics (QED) and electroweak (EW) corrections
to the Higgs decay process $H\to WW/ZZ\to4\ell$ were calculated in
Ref.\ \cite{Bredenstein:2006rh} and Ref.\ \cite{CarloniCalame:2006vr},
respectively. Recently, the NLO QCD correction to the Higgs boson
decays $H\to WW/ZZ\to4q$ with hadronic four-fermion final states
was calculated in Ref.\ \cite{Bredenstein:2006ha}. Since the higher
order corrections for Higgs production are dominated by the initial
state soft-gluon resummation effects, we focus our attention on the
RES effects in this work. It is worth mentioning that the NLO QED
corrections to the Higgs boson decay $H\to WW/ZZ\to4\ell$ have been
implemented in \textsc{ResBos}~\cite{Balazs:1997xd} program, and
the phenomenological study of the combined RES effects and the QED
correction will be presented elsewhere.%
}.

The paper is organized as follows. In Sec.\ \ref{sec:css}, we present
our analytical formalism of the CSS resummation. In Sec.\ \ref{sec:inclusive-xsec},
we present the inclusive cross section of the signal process for several
benchmark masses of the Higgs boson. In Sec.\ \ref{sec:hww}, we
study the process $gg\to H\to WW^{(*)}\to\ell^{+}\ell^{\prime-}\nu_{\ell}\bar{\nu}_{\ell^{\prime}}$
for $M_{H}=140\,$GeV at the Fermilab Tevatron and for $M_{H}=170\,$GeV
at the LHC. In Sec.\ \ref{sec:hzz}, we examine the process $gg\to H\to ZZ^{(*)}\to\ell^{+}\ell^{-}\ell^{\prime+}\ell^{\prime-}$
for $M_{H}=140\,$GeV and $200\,$GeV, respectively, and the process
$gg\to H\to ZZ\to\ell^{+}\ell^{-}\nu\bar{\nu}$ for $M_{H}=600\,$GeV,
at the LHC. Our conclusions are given in Sec.~\ref{sec:Conclusion}.

\section{Transverse momentum resummation formalism\label{sec:css}}

\begin{figure}
\includegraphics[clip,scale=0.5]{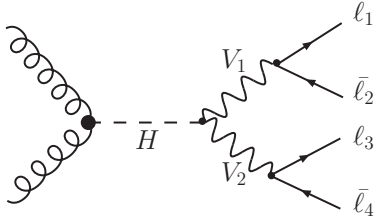}

\caption{Tree level Feynman diagram of process $gg\to H\to V_{1}(\to\ell_{1}\bar{\ell}_{2})V_{2}(\to\ell_{3}\bar{\ell}_{4})$.\label{fig:Feyn-daigr}}
\end{figure}

At the hadron colliders, the SM Higgs boson is mainly produced via
gluon-gluon fusion process through a heavy quark triangle loop diagram,
cf. Fig~\ref{fig:Feyn-daigr}, in which the effect of the triangle
loop is replaced by the effective $ggH$ coupling (denoted as the
bold dot). Taking advantage of the narrow width of the Higgs boson,
we can factorize the Higgs boson production from its sequential decay.
The resummation formula was already presented in Ref.\ \cite{Balazs:2000wv}.
Here, we list some of the relevant formulas as follows, for completeness:\begin{eqnarray}
 &  & \frac{d\sigma(h_{1}h_{2}\to H(\to VV\to\ell_{1}\ell_{2}\ell_{3}\ell_{4})X)}{dQ^{2}dQ_{T}^{2}dyd\phi_{H}d\Pi_{4}}\nonumber \\
 & = & \sigma_{0}(gg\to H)\frac{Q^{2}}{S}\frac{Q^{2}\Gamma_{H}/m_{H}}{(Q^{2}-m_{H}^{2})^{2}+(Q^{2}\Gamma_{H}/m_{H})^{2}}\nonumber \\
 & \times & \Biggl|\mathcal{M}(H\to V_{1}V_{2}\to\ell_{1}\ell_{2}\ell_{3}\ell_{4})\Biggr|^{2}\nonumber \\
 & \times & \Biggl\{\frac{1}{(2\pi)^{2}}\int d^{2}b\, e^{iQ_{T}\cdot b}\tilde{W}_{gg}(b_{*},Q,x_{1},x_{2},C_{1,2,3})\tilde{W}_{gg}^{NP}(b,Q,x_{1},x_{2})+Y(Q_{T},Q,x_{1},x_{2},C_{4})\Biggr\}\,\,\,\,\,\,\,\,\,\label{eq:resum}\end{eqnarray}
 where $Q$, $Q_{T}$, $y$, and $\phi_{H}$ are the invariant mass,
transverse momentum, rapidity, and azimuthal angle of the Higgs boson,
respectively, defined in the lab frame, and $d\Pi_{4}$ represents
the four-body phase space of the Higgs boson decay, defined in the
Collin-Soper frame~\cite{Collins:1977iv}. In Eq.~(\ref{eq:resum}),
$\left|\mathcal{M}(\cdots)\right|^{2}$ denotes the matrix element
square of the Higgs boson decay and reads as\begin{eqnarray*}
 &  & \Biggl|\mathcal{M}(H\to V_{1}V_{2}\to\ell_{1}\ell_{2}\ell_{3}\ell_{4})\Biggr|^{2}\\
 & = & 16\sqrt{2}G_{F}^{3}m_{V}^{8}\frac{1}{(q_{1}^{2}-m_{V}^{2})^{2}+m_{V}^{2}\Gamma_{V}^{2}}\frac{1}{(q_{2}^{2}-m_{V}^{2})^{2}+m_{V}^{2}\Gamma_{V}^{2}}\\
 & \times & \Biggl[C_{+}(p_{1}\cdot p_{3})(p_{2}\cdot p_{4})+C_{-}(p_{1}\cdot p_{4})(p_{2}\cdot p_{3})\Biggr],\end{eqnarray*}
 where $m_{V}$ is the vector boson mass, $q_{i}$($p_{i}$) denotes
the momentum%
\footnote{The direction of momentum $p_{i}$ is defined to be outgoing from
the mother particle.%
} of the vector boson $V_{i}$ (the lepton $\ell_{i}$), and $G_{F}$
is the Fermi coupling constant. Here,\[
C_{\pm}=\left(a_{12}^{2}+b_{12}^{2}\right)\left(a_{34}^{2}+b_{34}^{2}\right)\pm4a_{12}b_{12}a_{34}b_{34},\]
 where $a_{12}$ and $b_{12}$ respectively denote the vector and
axial vector components of the $V\ell_{1}\ell_{2}$ coupling, while
$a_{34}$ and $b_{34}$ are the ones for $V\ell_{3}\ell_{4}$. For
the $W$ boson, $m_{V}=m_{W}$, and \[
a=b=\sqrt{2},\]
 while for the $Z$ boson, $m_{V}=m_{Z}$, and \begin{eqnarray*}
a & =4\sin\theta_{W}^{2}-1, & \, b=-1\,\,\,{\rm for}\,\,\, Z\to\ell^{+}\ell^{-},\\
 & a=1,\, b=1 & {\rm for}\,\,\, Z\to\nu\bar{\nu},\end{eqnarray*}
 where $\theta_{W}$ is the weak mixing angle. In Eq.~(\ref{eq:resum}),
the function $\tilde{W}_{gg}$ sums over the soft gluon contributions
that grow as $Q_{T}^{-2}\times[1\,{\rm or}\,\,{\rm ln}(Q_{T}^{2}/Q^{2})]$
to all order in $\alpha_{S}$, which contains the singular part as
$Q_{T}\to0$. The contribution which is less singular than those included
in $\tilde{W}_{gg}$ is calculated order-by-order in $\alpha_{S}$
and is included in the $Y$ term. Therefore, we can obtain the NLO
results by expending the above resummation formula, i.e. Eq.~(\ref{eq:resum}),
to the $\alpha_{S}^{3}$ order. More details can be found in Ref.\ \cite{Balazs:1997xd}.
In our calculation, $\sigma_{0}$ includes the complete LO contribution
with finite quark mass effects~\cite{Wilczek:1977zn,Ellis:1979jy,Georgi:1977gs,Rizzo:1979mf}.
It has been shown~\cite{Kramer:1996iq} that this prescription approximates
well the exact NLO inclusive Higgs production rate.

For the numerical evaluation, we chose the following set of SM input
parameters~\cite{lepewwg:2003ih}:\begin{eqnarray*}
G_{F}=1.16637\times10^{-5}{\rm GeV}^{-2}, &  & \alpha=1/137.0359895,\\
m_{Z}=91.1875\,{\rm GeV}, &  & \alpha_{s}(m_{Z})=0.1186,\\
m_{e}=0.5109997\,{\rm MeV}, &  & m_{\mu}=0.105658389\,{\rm GeV}.\end{eqnarray*}
 Following Ref.~\cite{Degrassi:1997iy}, we derive the $W$ boson
mass as $m_{W}=80.385\,{\rm GeV}$. Thus, the square of the weak gauge
coupling is $g^{2}=4\sqrt{2}m_{W}^{2}G_{F}$. Including the $\mathcal{O}(\alpha_{s})$
QCD corrections to $W\to q\bar{q'}$, we obtain the $W$ boson width
as $\Gamma_{W}=2.093\,{\rm GeV}$ and the decay branching ratio of
${\rm Br}(W\to\ell\nu)=0.108$~\cite{Cao:2004yy}. In order to include
the effects of the higher order electroweak corrections, we also adapt
the effective Born approximation in the calculation of the $H\to ZZ\to4\,{\rm leptons}$
mode by replacing the $\sin^{2}\theta_{W}$ in the $Z\ell\ell$ coupling
by the effective $\sin^{2}\theta_{W}^{eff}=0.2314$, calculated at
the $m_{Z}$ scale.

\section{Inclusive cross sections\label{sec:inclusive-xsec}}

For the mass of the Higgs boson being within the intermediate mass
range, it will principally decay into two vector bosons which sequentially
decay into either lepton or quark pairs. Leptons are the objects which
can be easily identified in the final state, so the di-lepton decay
mode is regarded as the {}``golden channel'' due to its clean signature
and well-known background. The drawback is that the di-lepton mode
suffers from the small decay branching ratio for the vector boson
decay ($V\to\ell\bar{\ell}$). For example, the branching ratio of
$Z\to\ell^{+}\ell^{-}$ is only about $3.4\%$. Due to the huge QCD
backgrounds, the purely hadronic decay modes are not as useful for
detecting the Higgs boson.

In this paper, we focus on the purely leptonic decays of the vector
bosons in the $H\to WW^{(*)}$ and $H\to ZZ^{(*)}$ modes. To cover
the intermediate mass range, we consider the following benchmark cases:
(i) $H\to WW^{(*)}\to\ell^{+}\ell^{\prime-}\nu_{\ell}\bar{\nu}_{\ell^{\prime}}$
($\ell,\ell^{\prime}=e$\ or\ $\mu$) for $M_{H}=140\,$GeV at the
Femilab Tevatron Run 2 (a 1.96 TeV $p\bar{p}$ collider), and for
$M_{H}=170\,$GeV at the LHC (a 14 TeV $pp$ collider); (ii) $H\to ZZ^{(*)}\to\ell^{+}\ell^{-}\ell^{\prime+}\ell^{\prime-}$
($\ell,\ell^{\prime}=e$\ or\ $\mu$) for $M_{H}=140$ and $200\,$GeV
at the LHC; (iii) $H\to ZZ\to\ell^{+}\ell^{-}\nu\bar{\nu}$ for $M_{H}=600\,$GeV
at the LHC, where $\ell=e$ or $\mu$, and $\nu=\nu_{e}$, $\nu_{\mu}$
or $\nu_{\tau}$. All the numerical results are calculated by using
ResBos~\cite{Balazs:1997xd}. We adapt CTEQ6.1L parton distribution
function in the LO calculation and CTEQ6.1M parton distribution function~\cite{Pumplin:2002vw}
in the NLO and RES calculations. The renormalization scale ($\mu_{R}$)
and factorization scale ($\mu_{F}$) are chosen to be the Higgs boson
mass in our calculations, i.e. $\mu_{R}=\mu_{F}=M_{H}$.

\begin{table}

\caption{Inclusive cross sections of $gg\to H\to VV\to4\ell$ at the Tevatron
Run 2 and the LHC in the unit of fb, i.e. $\sigma(gg\to H)\times Br(H\to VV)\times Br(V\to\ell_{1}\ell_{2})\times Br(V\to\ell_{3}\ell_{4})$
for various Higgs boson masses. Here, $\ell$ and $\ell^{\prime}$
denote either $e$ or $\mu$. ~\label{tab:Total-cross-section}}

\begin{tabular}{c|c|c|c|c|c}
\hline 
&
\multicolumn{2}{c|}{$WW^{(*)}\to\ell^{+}\ell^{\prime-}\nu_{\ell}\bar{\nu}_{\ell^{\prime}}$}&
\multicolumn{2}{c|}{$ZZ^{(*)}\to\ell^{+}\ell^{-}\ell^{\prime+}\ell^{\prime-}$}&
$ZZ\to\ell^{+}\ell^{-}\nu\bar{\nu}(=\underset{i=e,\mu,\tau}{\sum}\nu_{i}\bar{\nu}_{i})$\tabularnewline
\hline 
$M_{H}$&
$140\,$GeV &
$170\,$GeV &
$140\,$GeV&
$200\,$GeV &
$600\,$GeV \tabularnewline
&
Tevatron&
LHC&
LHC&
LHC&
LHC\tabularnewline
\hline 
RES&
$13.1$&
$891.1$&
$11.0$&
$17.7$&
$6.3$\tabularnewline
NLO&
$11.5$&
$848.9$&
$10.5$&
$16.4$&
$5.6$\tabularnewline
LO&
$4.0$&
$405.3$&
$5.1$&
$8.0$&
$2.4$\tabularnewline
\hline
\end{tabular}
\end{table}

\begin{figure}
\includegraphics[scale=0.4]{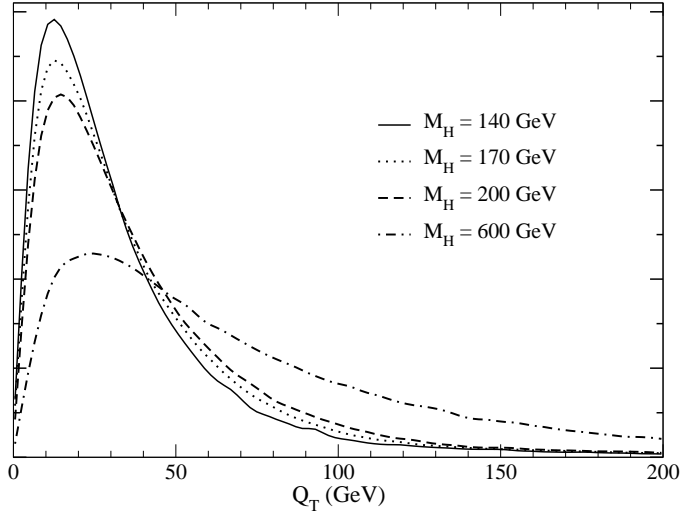}

\caption{Normalized distributions of transverse momentum of Higgs boson predicted
by RES calculation at the LHC.~\label{fig:QT_H_all}}
\end{figure}

\begin{figure}
\includegraphics[clip,scale=0.7]{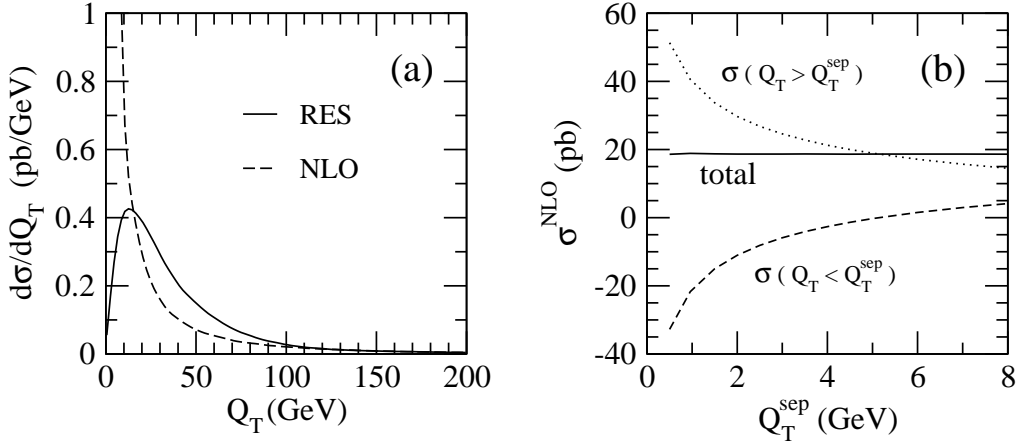}

\caption{(a) Distribution of transverse momentum of Higgs boson, and (b) NLO
total production cross section of Higgs boson via gluon gluon fusion
as $M_{H}=170\,$GeV at the LHC.~\label{fig:pt_H}}
\end{figure}

The inclusive cross sections for those benchmark masses of the Higgs
boson are summarized in Table~\ref{tab:Total-cross-section} where
different searching channels are considered. For comparison, we show
the $Q_{T}$ distributions calculated by using the RES and NLO calculations
in Fig.~\ref{fig:pt_H}(a). The RES calculation is similar to that
presented in Ref.~\cite{Balazs:2000wv,Berger:2002ut} with the known
$A$ and $B$~ \cite{Kauffman:1991jt,Kauffman:1991cx,Yuan:1991we,deFlorian:2000pr,deFlorian:2001zd},
but with $A_{g}^{(3)}$ included, where~\cite{Vogt:2004mw}\begin{eqnarray}
A_{g}^{(3)} & = & \frac{C_{A}C_{F}N_{f}}{2}(\zeta(3)-\frac{55}{48})-\frac{C_{A}N_{f}^{2}}{108}+C_{A}^{3}(\frac{11\zeta(3)}{24}+\frac{11\pi^{4}}{720}-\frac{67\pi^{2}}{216}+\frac{245}{96})\nonumber \\
 &  & +C_{A}^{2}N_{f}(-\frac{7\zeta(3)}{12}+\frac{5\pi^{2}}{108}-\frac{209}{432})\,,\label{eq:A3}\end{eqnarray}
 where $C_{A}=3$, $C_{F}=4/3$, $N_{f}=5$ and the Riemann constant
$\zeta(3)=1.202...$ . We also use the modified parton momentum fractions
$x_{1}$ and $x_{2}$ to take into account the kinematic corrections
due to the emitted soft gluons~\cite{Balazs:2000wv}, with $x_{1}=m_{T}e^{y}/\sqrt{S}$
and $x_{2}=m_{T}e^{-y}/\sqrt{S}$, where $m_{T}=\sqrt{Q_{T}^{2}+Q^{2}}$
and $\sqrt{S}$ is the center-of-mass energy of the hadron collider.
We also adopt the matching procedure described in the Ref.~\cite{Balazs:1997xd}
and the non-perturbation contribution $\tilde{W}^{NP}$ of BLNY form
in the Ref.~\cite{Landry:2002ix}. In Fig.~\ref{fig:QT_H_all},
we show the transverse momentum distributions of Higgs boson predicted
by RES calculation at the LHC. As we see that the peak position is
shifted to larger $Q_{T}$ region and the shape becomes broader when
the mass of Higgs becomes heavier. 

It is clear that the prediction of NLO calculation blows up in the
$Q_{T}\to0$ region and the RES effects have to be included to make
a reliable prediction on event shape distributions. In the NLO calculation,
it is ambiguous to treat the singularity of the $Q_{T}$ distribution
near $Q_{T}=0$, see the dashed curve in Fig.\ \ref{fig:pt_H}(a).
Before presenting our numerical results, we shall explain how we deal
with the singularity in the NLO calculation when $Q_{T}\sim0$. In
ResBos, we divide the $Q_{T}$ phase space with a separation scale
$Q_{T}^{sep}$. We calculate the $Q_{T}$ singular part of real emission
and virtual correction diagrams analytically and integrate the sum
of these two parts up to $Q_{T}^{sep}$. By this procedure, it yields
a finite NLO cross section, for integrating $Q_{T}$ from 0 up to
$Q_{T}^{sep}$, which is put into the $Q_{T}=0$ bin of the NLO $Q_{T}$
distribution (for bin width larger than $Q_{T}^{sep}$). Since the
separation scale $Q_{T}^{sep}$ is introduced in the theoretical calculation
for technical reasons only and is not a physical observable, the sum
of both contributions from $Q_{T}>Q_{T}^{sep}$ and $Q_{T}<Q_{T}^{sep}$
should not depend on $Q_{T}^{sep}$. As shown in Fig.~\ref{fig:pt_H}(b),
the NLO total cross section indeed does not depend on the choice of
$Q_{T}^{sep}$ as long as it is not too large. We refer the readers
to the Sec. 3 and the Appendix of Ref.~\cite{Balazs:1997xd} for
more details. In this study, we choose $Q_{T}^{sep}=0.96\,$GeV in
our numerical calculations.

\begin{figure}
\includegraphics[scale=0.6]{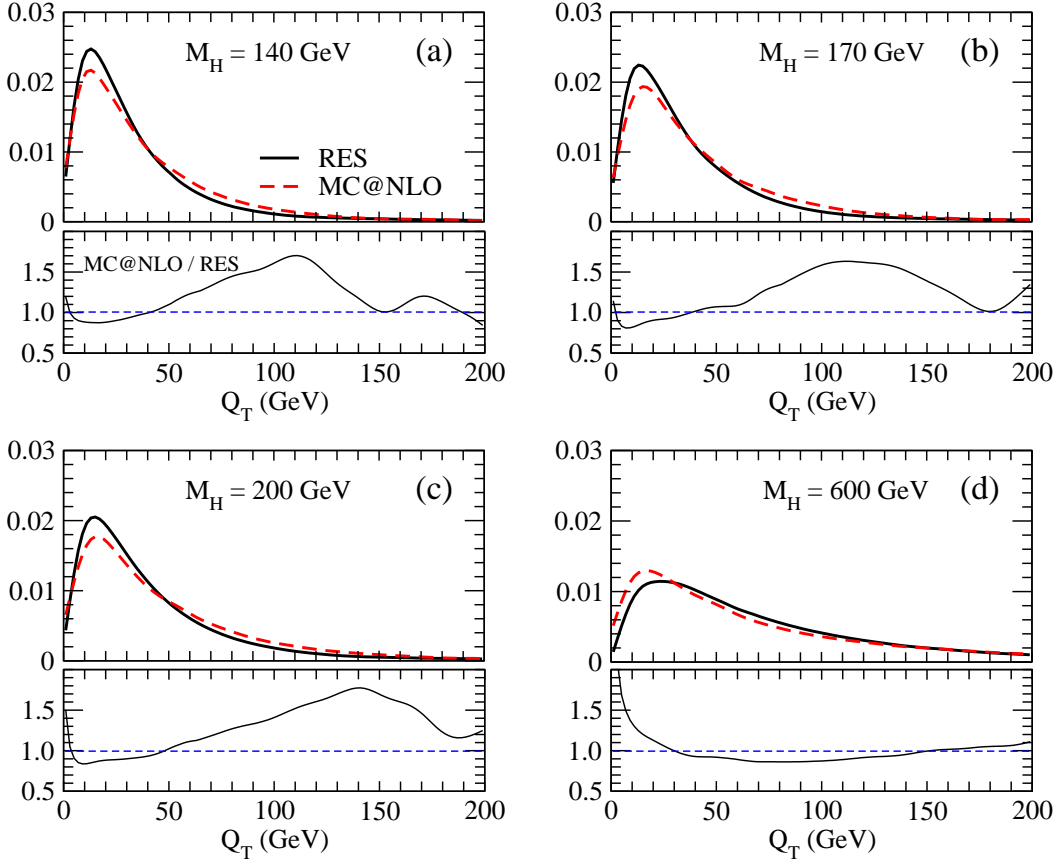}

\caption{Comparison of the $Q_{T}$ distributions between ResBos and MC@NLO.\label{fig:resbos.vs.mcnlo}}
\end{figure}

\textcolor{black}{As mentioned in the Introduction, MC@NLO, which
matches NLO calculations and parton showering Monte Carlo event generators,
not only predicts a reliable $Q_{T}$ of the Higgs boson but also
includes spin correlations among the Higgs decay products. Therefore
it is interesting to compare the $Q_{T}$ predictions between MC@NLO
and RES calculations. In order to compare the differences in shape
more precisely, we show the $Q_{T}$ distributions predicted by MC@NLO
and ResBos in Fig.~\ref{fig:resbos.vs.mcnlo} for $M_{H}=140(\,170,\,200,\,600)\,{\rm GeV}$.
All distributions are normalized by the total cross sections for the
corresponding Higgs boson masses. The bottom part of each $Q_{T}$
distribution plot presents the ratio between MC@NLO and ResBos. We
note that for a light Higgs boson the distributions are consistent
in the peak region\ \cite{Huston:2004yp,Dobbs:2004bu}, where the
difference is about $10\,\%$, but they are quite different in the
large $Q_{T}$ region, say $Q_{T}\gtrsim100\,{\rm GeV}$. For a heavy
Higgs boson, e.g. $M_{H}=600\,{\rm GeV}$, these two distributions
are very different in the small $Q_{T}$ region, and MC@NLO tends
to populate more events in the small $Q_{T}$ region, as compared
to ResBos. Since the Higgs boson is a scalar, the distributions of
Higgs boson decay products just depend upon the Higgs boson's kinematics.
Therefore, the difference in the $Q_{T}$ distribution predictions
between MC@NLO and ResBos may prove to be crucial for the precision
measurements of the Higgs boson's properties. A further detailed study
of the impact of the $Q_{T}$ difference on the Higgs boson search
is in order and will be presented elsewhere.}

\section{Phenomenological study of the $H\to WW$ mode\label{sec:hww}}

In the search for SM-like Higgs boson via $H\rightarrow WW^{(*)}$
mode, two scenarios of $W$ boson decay were considered in the literature~\cite{Dittmar:1996ss,Han:1998ma,Han:1998sp}:
one is that both $W$ bosons decay leptonically, another is that one
$W$ boson decays leptonically and another $W$ boson decays hadronically.
Throughout this paper, we only concentrate on the di-lepton decay
mode, i.e. $H\rightarrow WW^{(*)}\rightarrow\ell^{+}\ell^{\prime-}\nu_{\ell}\bar{\nu}_{\ell^{\prime}}$,
at the Tevatron and the LHC. The collider signature, therefore, is
two isolated opposite-sign charged leptons plus large missing transverse
energy ($\met$) which originates from the two neutrinos. In this
section, we first examine the RES effects on various kinematics distributions,
and then show the RES effects on the Higgs mass measurement. Finally,
we study the RES effects on the acceptances of the kinematics cuts
suggested in the literature for Higgs search.

\subsection{Basic kinematics distributions}

\begin{figure}
\includegraphics[scale=0.7]{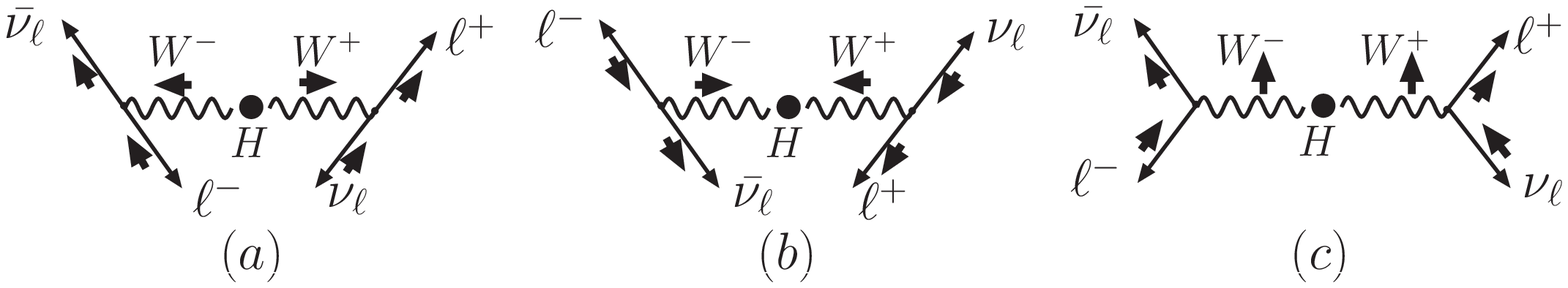}

\caption{Kinematic configurations of Higgs decay ($H\to WW\to\ell^{+}\ell^{-}\nu\bar{\nu}$)
in the rest frame of $H$: (a) $H\to W_{+}^{+}W_{+}^{-}$, (b) $H\to W_{-}^{+}W_{-}^{-}$
and (c) $H\to W_{0}^{+}W_{0}^{+}$. Here, $+(-,\,0$) denotes the
right-handed (left-handed, longitudinal) polarization state of the
$W$ boson. The long arrows denote the moving directions of the final-state
leptons. The short bold arrows denote the particles' spin directions.\label{fig:spin}}
\end{figure}

\begin{figure}
\includegraphics[clip,scale=0.6]{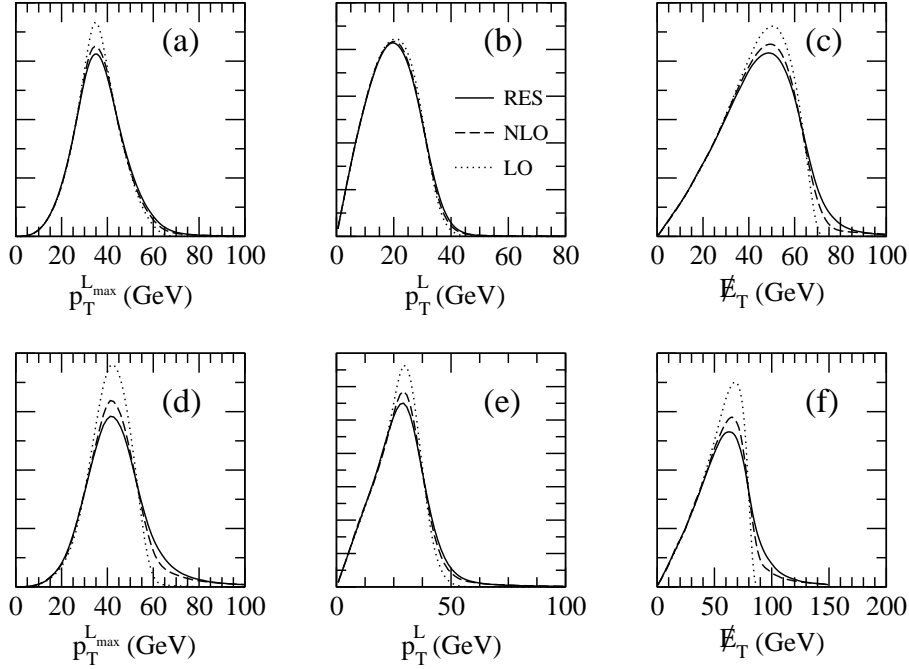}

\caption{Normalized distributions of the leading transverse momentum $p_{T}^{L_{max}}$,
softer transverse momentum $p_{T}^{L}$ of the leptons, and the missing
energy $\met$ in $gg\to H\to WW\to\ell^{+}\ell^{\prime-}\nu_{\ell}\bar{\nu}_{\ell^{\prime}}$.
The panels (a) to (c) are for $M_{H}=140\,{\rm GeV}$ at the Tevatron,
and (d) to (f) are for $M_{H}=170\,{\rm GeV}$ at the LHC.~\label{fig:nocut}}
\end{figure}

\begin{figure}
\includegraphics[clip,scale=0.7]{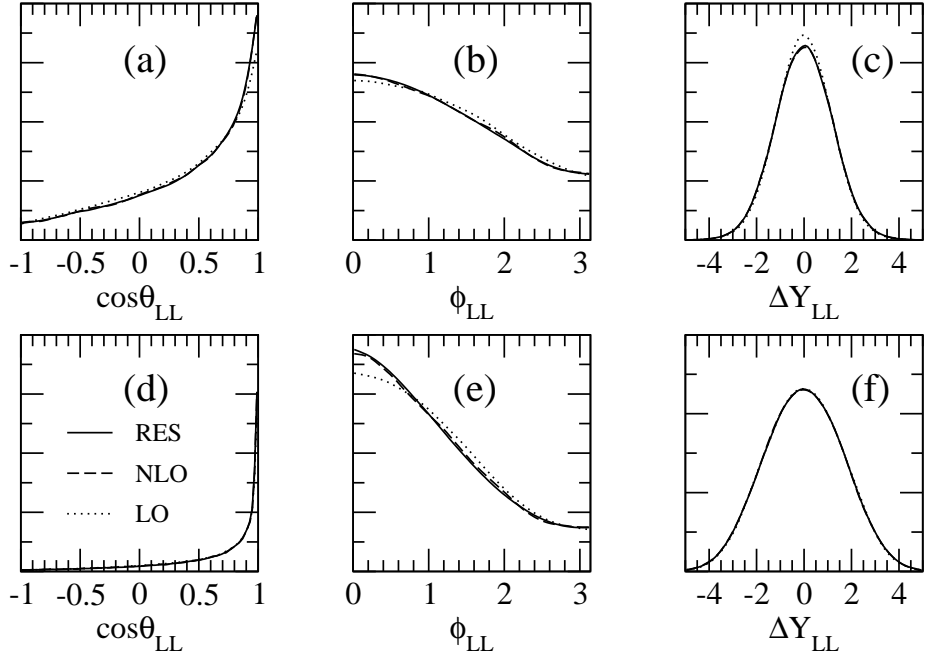}

\caption{Normalized distributions of $\cos\,\theta_{LL}$, $\phi_{LL}$ and
$\Delta Y_{LL}$ in $gg\to H\to WW\to\ell^{+}\ell^{\prime-}\nu_{\ell}\bar{\nu}_{\ell^{\prime}}$:
The panels~(a) to (c) are for $M_{H}=140\,{\rm GeV}$ at the Tevatron
and (d) to (f) are for $M_{H}=170\,{\rm GeV}$ at the LHC.~\label{fig:w_angle}}
\end{figure}

For a heavy Higgs boson, the two vector bosons, which are generated
from the spin-0 Higgs boson decay, are predominantly longitudinally
polarized, while the longitudinal and transverse polarization states
are democratically populated when the Higgs boson mass is near the
threshold for decaying into the vector boson pair\ \cite{Kane:1989vv,Barger:1993wt}.
When ${140\,{\rm GeV}\leq M}_{H}\leq170\,{\rm GeV}$, the transverse
polarization modes contribute largely. The two charged leptons in
the final state have different kinematics because of the conservation
of angular momentum, cf. Fig.\ \ref{fig:spin}, therefore, one charged
lepton is largely boosted and its momentum becomes harder while another
becomes softer. Making use of these differences, one can impose asymmetric
transverse momentum ($p_{T}$) cuts on the two charged leptons to
suppress the background. On the event-by-event basis, we arrange the
two charged leptons in the order of transverse momentum: $p_{T}^{L_{max}}$
denotes the larger $p_{T}$ between the two charged leptons while
$p_{T}^{L}$ is the smaller one. Fig.~\ref{fig:nocut} shows the
distributions of $p_{T}^{L_{max}}$, $p_{T}^{L}$ and missing energy
($\met$) for $M_{H}=140\,{\rm GeV}$ at the Tevatron (first row)
and for $M_{H}=170\,{\rm GeV}$ at the LHC (second row). Furthermore,
in Fig.~\ref{fig:w_angle} we show the distributions of $\cos\theta_{LL}$,
$\phi_{LL}$ and $\Delta Y_{LL}$ without imposing any kinematics
cut, where $\cos\theta_{LL}$ is the cosine of the opening angle between
the two charged leptons, $\phi_{LL}$ is the azimuthal angle difference
between the two charged leptons on the transverse plane, and $\Delta Y_{LL}$
is the rapidity difference of two charged leptons in the lab frame.
Since we are mainly interested in the shapes of the kinematics distributions,
the curves shown in the figures are all normalized by the corresponding
total cross sections. The solid curves present the distributions including
the RES effects, the dashed and dotted curves present the distributions
calculated at the NLO and LO, respectively.

We note that the $p_{T}$ distributions of the charged leptons and
the missing energy distributions are modified largely by the RES effects.
This can be understood as follows. The two charged leptons prefer
to move in the same direction due to the spin correlation among the
decay products of the Higgs boson, cf. the distributions of $\cos\theta_{LL}$
in Figs.~\ref{fig:w_angle}(a)~and~(d). Hence, one can approximately
treat the Higgs boson decay as {}``two-body'' decay, i.e. decaying
into two clusters as $H\to\left(\ell^{+}\ell^{\prime-}\right)\left(\nu_{\ell}\bar{\nu}_{\ell^{\prime}}\right)$.
This is in analogy to the $W$ boson production and decay in the Drell-Yan
process, $u\bar{d}\to W^{+}\to\ell^{+}\nu$, which has been shown
in Ref.~\cite{Cao:2004yy} that the transverse momentum of lepton
($p_{T}^{\ell}$) is very sensitive to the transverse momentum of
the $W$ boson. The same sensitivity also applies to $\met$. As shown
in Figs.~\ref{fig:nocut}(c)~and~(f), the clear Jacobian peak of
the $\met$ distribution around $M_{H}/2$ in the LO calculation is
smeared in the NLO and RES calculations. Furthermore, the $\met$
distribution in the NLO and RES calculations has a long tail due to
the non-zero transverse momentum of the Higgs boson. Since the RES
calculation includes the effects from multiple soft-gluon radiation,
the $\met$ distribution near the Jacobian peak is further smeared
in the RES calculation as compared to the NLO calculation. When $M_{H}=140\,$GeV,
only one $W$ boson is on-shell and the two charged leptons do not
move as close as they do in the case of $M_{H}=170\,$GeV (in which
case, both $W$ bosons are on-shell). However the parallel configuration
is still preferred.

The dominant backgrounds of the $H\to WW^{(*)}$ mode are from the
$W$ boson pair production and top quark pair production. The latter,
as the reducible background, can be suppressed with suitable cuts
such as jet-veto, but the former, as the irreducible background, still
remains even after imposing the basic kinematic cuts. In order to
reduce this intrinsic background, one needs to take advantage of the
characteristic spin correlations of the charged leptons in the $H\to WW^{(*)}\to\ell^{+}\ell^{\prime-}\nu_{\ell}\bar{\nu}_{\ell^{\prime}}$
decay. For example, the distribution of the difference in azimuthal
angles of the charged leptons peaks at smaller value (cf. Figs.~\ref{fig:w_angle}(b)\ and\ (e))
for the signal than that for the $WW$ continuum production background~\cite{Dittmar:1996ss,Han:1998sp}.
We note that the RES effects do not affect the $\cos\theta_{LL}$
and $\phi_{LL}$ distributions very much, as shown in Figs.~\ref{fig:w_angle}(a),
(b), (d) and (e).

\begin{figure}
\includegraphics[clip,scale=0.6]{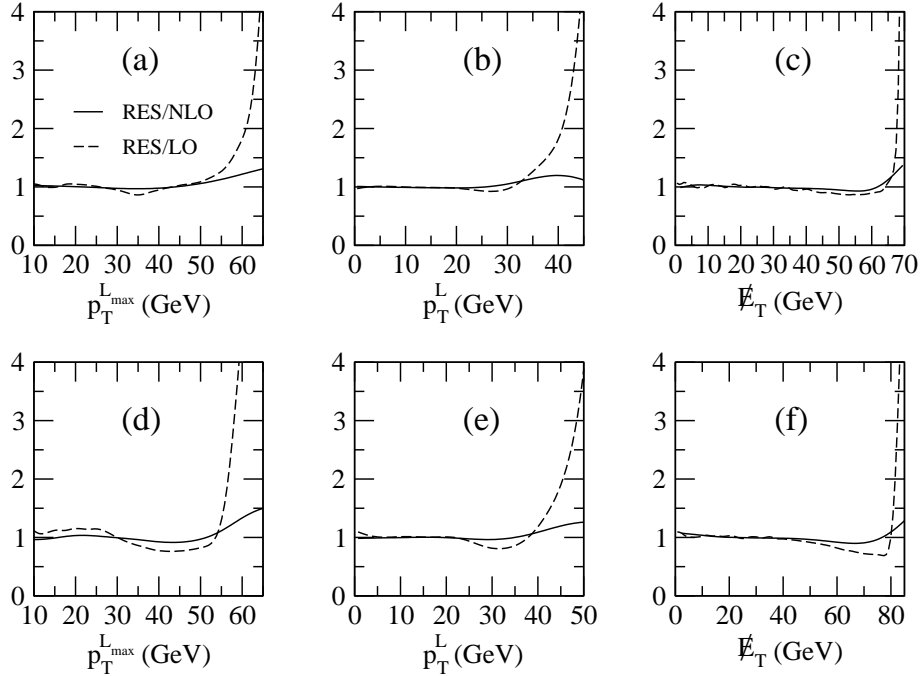}

\caption{Ratio of the Resummation contribution to NLO and LO contributions
in $gg\to H\to WW^{(*)}\to\ell^{+}\ell^{\prime-}\nu_{\ell}\bar{\nu}_{\ell^{\prime}}$.
The panels (a) to (c) are for $M_{H}=140\,{\rm GeV}$ at the Tevatron
while (d) to (f) are for $M_{H}=170\,{\rm GeV}$ at the LHC.\label{fig:w-ratio}}
\end{figure}

To closely examine the difference in their predictions, we also present
the ratio of the RES contribution to the NLO and LO contributions
in Fig.~\ref{fig:w-ratio}. We note that the ratio is about one below
the peak regions of $p_{T}^{L_{max}}$, $p_{T}^{L}$ and $\met$,
and becomes larger than one above the peak region, where both the
LO and NLO contributions drop faster than the RES contribution does,
which is consistent with the results shown in Fig.~\ref{fig:nocut}.
This uneven behavior indicates that one cannot simply use the leading
order kinematics with the constant $K$-factor included to mimic the
higher order quantum corrections. We should stress that even though
the NLO and RES calculations include the same contributions of the
hard gluon radiation from initial states, the effects of the multiple
soft-gluon radiation could cause more than \textcolor{black}{$25\%$}
difference between RES and NLO predictions in the large $p_{T}$ and
$\met$ region.

\subsection{Higgs mass measurement}

\begin{figure}
\includegraphics[clip,scale=0.6]{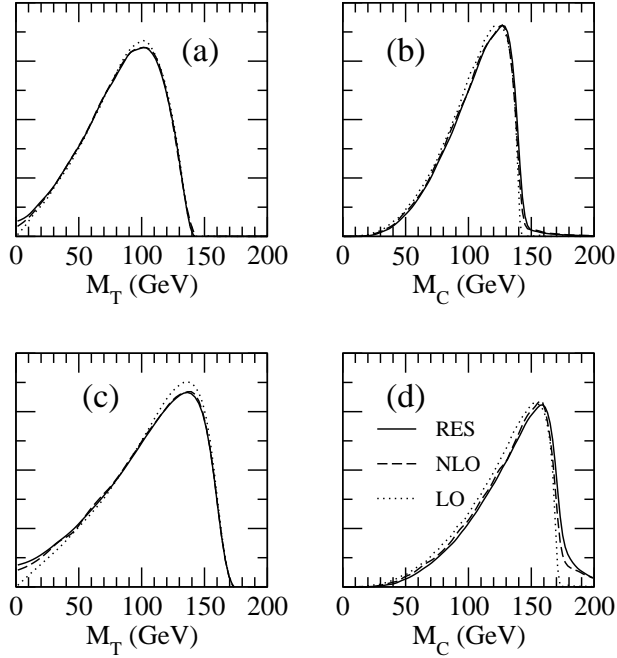}

\caption{Normalized distributions of the transverse mass $M_{T}$ and the
cluster mass $M_{C}$ in $gg\to H\to WW^{(*)}\to\ell^{+}\ell^{\prime-}\nu_{\ell}\bar{\nu}_{\ell^{\prime}}$
: (a) and (b) are for $M_{H}=140\,{\rm GeV}$ at the Fermilab Tevatron
while (c) and (d) are for $M_{H}=170\,{\rm GeV}$ at the LHC.~\label{fig:w-mt-mc}}
\end{figure}

In order to identify the signal events clearly, it is crucial to reconstruct
the invariant mass of the Higgs boson. Unfortunately, one cannot directly
reconstruct the $M_{H}$ distribution in the $H\to WW$ mode due to
the two neutrinos in the final state. Instead, both the transverse
mass $M_{T}$ and the cluster transverse mass $M_{C}$~\cite{Barger:1987re},
defined as \begin{eqnarray}
M_{T} & = & \sqrt{2p_{T}^{LL}\met(1-\cos\Delta\phi(p_{T}^{LL},\met))},\nonumber \\
M_{C} & = & \sqrt{p_{T}^{LL^{2}}+m_{LL}^{2}}+\met,\label{eq:mt-mc}\end{eqnarray}
 yield a broad peak near $M_{H}$. In Eq.~(\ref{eq:mt-mc}), $p_{T}^{LL}$
($m_{LL}$) denotes the transverse momentum (invariant mass) of the
two charged lepton system, and $\Delta\phi(p_{T}^{LL},\met)$ is the
difference in azimuthal angles between $p_{T}^{LL}$ and $\met$ on
the transverse plane. We note that the upper endpoint of $M_{T}$
distribution can clearly reflect the mass of Higgs boson, cf. Figs.\ \ref{fig:w-mt-mc}(a)\ and\ (c).
$M_{T}$ is insensitive to $Q_{T}$ because it depends on $Q_{T}$
in the second order, cf. Eq.\ (\ref{eq:mt-mc}). Therefore, the position
of the endpoint is only subject to $M_{H}$ and $\Gamma_{H}$. The
latter effects can be safely ignored because $\Gamma_{H}$ is very
small (less than about $1.5\,{\rm GeV}$), for the Higgs boson mass
less than $200\,{\rm GeV}$. The cluster transverse mass $M_{C}$
also exhibits a clear Jacobian peak with a clear edge at $M_{H}$,
cf. Figs.~\ref{fig:w-mt-mc}(b)\ and\ (d). But both the line shape
and the Jacobian peak of $M_{C}$ distribution are modified by the
RES effects because $M_{C}$ is directly related to $\met$ which
depends on $Q_{T}$ in the first order. We suggest that one should
use $M_{T}$ to extract the mass of Higgs in $H\to WW^{(*)}\to\ell^{+}\ell^{\prime-}\nu_{\ell}\bar{\nu}_{\ell^{\prime}}$
mode because the upper endpoint of the $M_{T}$ distribution is insensitive
to high order corrections.

\subsection{Acceptance study}

In order to separate the signal from its copious backgrounds, one
needs to impose optimal cuts to suppress backgrounds and enhance the
signal to background ratio ($S/B$ ) simultaneously. The selection
of the optimal cuts highly depends on how well we understand the kinematics
of the signal and background processes. As shown above, the RES effects
modify the distributions of transverse momentum of the charged leptons
and the missing energy largely, therefore, it is important to study
the RES effects on the acceptances of the kinematics cuts. Here, we
impose a set of kinematics cuts used by experimental colleagues in
Refs.~\cite{atlas:1999,Abazov:2005un}. The corresponding acceptances
are summarized in Table~\ref{tab:w-accept}.

\begin{itemize}
\item For the search for a 140~GeV Higgs boson at the Tevatron, we impose
the following \emph{basic cuts}:\begin{eqnarray}
p_{T}^{L_{max}}>15\,{\rm GeV} & , & p_{T}^{L}>10\, GeV,\nonumber \\
|Y_{L}|<2.0 & , & \met>20\,{\rm GeV},\label{eq:cut1-140}\end{eqnarray}
 and the \emph{optimal cuts} as follows: \begin{eqnarray}
m_{LL}<\frac{M_{H}}{2} & , & \frac{M_{H}}{2}<M_{T}<M_{H}-10\,{\rm GeV}\nonumber \\
\phi_{LL}<2.0\,{\rm rad} & , & \frac{M_{H}}{2}+20\, GeV<H_{T}<M_{H}\label{eq:cut2-140}\end{eqnarray}
 where $Y_{L}$ denotes the rapidity of charged lepton, and $H_{T}$
denotes the scalar sum of the transverse momenta of final state particles,
i.e. $H_{T}\equiv|p_{T}^{e}|+|p_{T}^{\mu}|+|\met|$. The overall efficiency
of the cuts is about $68\%$ , $69\%$ and $70\%$ after imposing
the basic cuts (Eq.~(\ref{eq:cut1-140})) for RES, NLO and LO calculations,
respectively, and about $44\%$ for both RES and NLO calculations
and $46\%$ for LO calculation after imposing the optimal cuts (Eq.~(\ref{eq:cut2-140})). 
\item For the search of a 170~GeV Higgs boson at the LHC, we require the
following \emph{basic cuts}:\begin{eqnarray}
p_{T}^{L_{max}}>20\,{\rm GeV} & , & p_{T}^{L}>10\,{\rm GeV},\nonumber \\
\left|Y_{L}\right|<2.5 & , & \met>40\,{\rm GeV},\label{eq:w170-cut1}\end{eqnarray}
 and the \emph{optimal cuts}:\begin{eqnarray}
m_{LL}<80.0\,{\rm GeV} & , & M_{H}-30.0\,{\rm GeV}<M_{T}<M_{H}\,,\nonumber \\
\phi_{LL}<1.0\,{\rm rad} & , & \theta_{LL}<0.9\,{\rm rad}\,\,,\,\,\,\,\,\left|\Delta Y_{LL}\right|<1.5\,\,,\label{eq:w170-cut2}\end{eqnarray}
 The cut efficiency is about $61\%$ for both RES and NLO calculations,
but about $63\%$ for LO contribution after imposing the basic cut
(Eq.~(\ref{eq:w170-cut1})). After imposing the optimal cuts (Eq.~(\ref{eq:w170-cut2})),
the acceptances of RES and NLO are about $19\%$, while LO is $20\%$. 
\end{itemize}
\begin{table}

\caption{Acceptance of $gg\to H\to WW^{(*)}\to\ell^{+}\ell^{\prime-}\nu_{\ell}\bar{\nu}_{\ell^{\prime}}$
events after imposing the basic cuts and the optimal cuts for $M_{H}=140\,{\rm GeV}$
at the Tevatron and $M_{H}=170\,{\rm GeV}$ at the LHC.\label{tab:w-accept}}

\begin{tabular}{c|c|c|c|c}
\hline 
&
\multicolumn{2}{c|}{$M_{H}=140\,{\rm GeV}$}&
\multicolumn{2}{c}{$M_{H}=170\,{\rm GeV}$}\tabularnewline
\hline
\hline 
&
basic (Eq.~(\ref{eq:cut1-140}))&
optimal (Eq.~(\ref{eq:cut2-140}))&
basic (Eq.~(\ref{eq:w170-cut1}))&
optimal (Eq.~(\ref{eq:w170-cut2}))\tabularnewline
\hline 
RES&
0.68&
0.44&
0.61&
0.19\tabularnewline
NLO&
0.69&
0.44&
0.61&
0.19\tabularnewline
LO&
0.70&
0.46&
0.63&
0.20\tabularnewline
\hline
\end{tabular}
\end{table}

\section{Phenomenological study of the $H\to ZZ$ mode\label{sec:hzz}}

In the search for the SM Higgs boson, the $H\to ZZ^{(*)}\to\ell^{+}\ell^{-}\ell^{\prime+}\ell^{\prime-}$
is an important discovery channel for a wide range of Higgs boson
mass. The appearance of four charged leptons with large transverse
momenta is an attractive experimental signature. This so-called {}``gold-plated''
mode provides not only a clean signature to verify the existence of
the Higgs boson but also an excellent process to explore its spin
and $CP$ properties~\cite{Buszello:2002uu}. In this section, we
study three mass values of $M_{H}$ ($140,\,200$ and $600\,{\rm GeV}$)
at the LHC. For $M_{H}=140\,$GeV and $200\,$GeV, we require the
two $Z$ bosons both decay into charged leptons; for $M_{H}=600\,$GeV,
we require one $Z$ boson decays into a charged lepton pair and another
$Z$ boson decays into a neutrino pair, i.e. $\ell^{+}\ell^{-}\nu\bar{\nu}$.
In this section we first study the RES effects on various kinematics
distributions and then examine the RES effects on the acceptances
of the kinematics cuts.

\subsection{$gg\to H\to ZZ^{(*)}\to\ell^{+}\ell^{-}\ell^{\prime+}\ell^{\prime-}$}

\begin{figure}
\includegraphics[clip,scale=0.6]{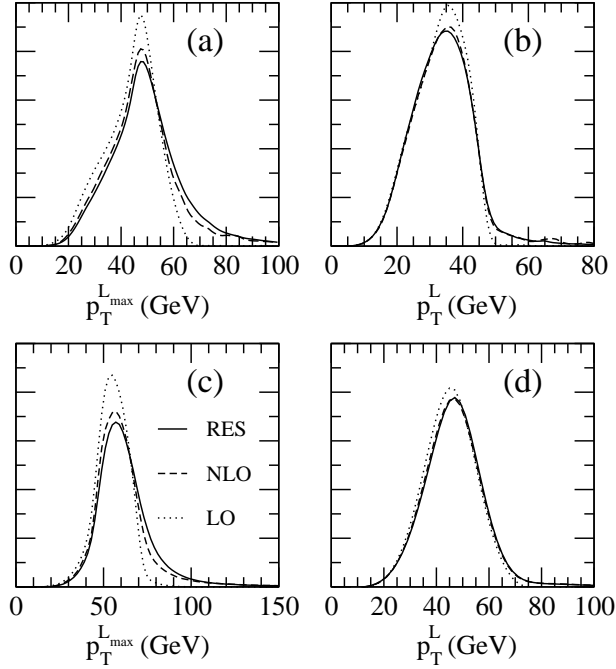}

\caption{Normalized distributions of $p_{T}^{L_{max}}$ and $p_{T}^{L}$ in
$gg\to H\to ZZ\to\ell^{+}\ell^{-}\ell^{\prime+}\ell^{\prime-}$: (a)
and (b) are for $M_{H}=140\,{\rm GeV}$; (c) and (d) are for $M_{H}=200\,{\rm GeV}$
at the LHC.\label{fig:z-pt}}
\end{figure}

Similar to the $H\to WW^{(*)}$ mode, we also arrange the four charged
leptons of the $H\to ZZ^{(*)}$ mode in the order of transverse momentum.
We denote $p_{T}^{L_{max}}$ as the largest $p_{T}$ of the four charged
leptons while $p_{T}^{L}$ the second leading $p_{T}$. In Fig.~\ref{fig:z-pt},
we show the distributions of $p_{T}^{L_{max}}$ and $p_{T}^{L}$ for
$M_{H}=140$ and $200\,$GeV, respectively. Due to the similar kinematics
discussed in the $H\to WW^{(*)}$ mode, the shapes of the distributions
of $p_{T}^{L_{max}}$ and $p_{T}^{L}$ are changed significantly by
the RES effects. The typical feature is that the RES effects shift
the $p_{T}$ of the charged lepton to the larger $p_{T}$ region and,
therefore, increase the acceptances of the kinematics cuts. The numerical
results will be shown later.

\begin{figure}
\includegraphics[clip,scale=0.6]{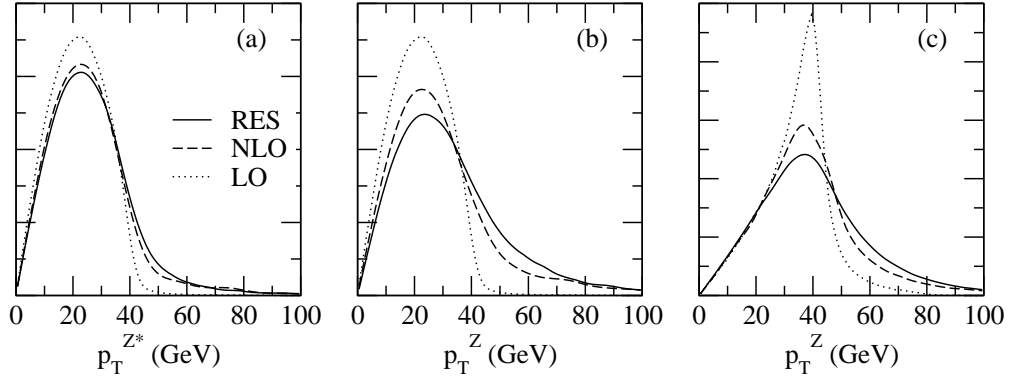}

\caption{Normalized transverse momentum of $Z$ boson in $gg\to H\to ZZ^{(*)}\to\ell^{+}\ell^{-}\ell^{\prime+}\ell^{\prime-}$
at the LHC: (a) is the $p_{T}$ distributions of off-shell $Z$ boson
for $M_{H}=140\,{\rm GeV}$, (b) is the $p_{T}$ distributions of
on-shell $Z$ boson for $M_{H}=140\,{\rm GeV}$ and (c) is the $p_{T}$
distributions of on-shell $Z$ boson for $M_{H}=200\,{\rm GeV}$.\label{fig:zrec}}
\end{figure}

Although one can measure the Higgs boson mass by reconstructing the
invariant mass of the four charged leptons, one still needs to reconstruct
the $Z$ bosons in order to suppress the backgrounds. The reconstruction
of the $Z$ boson depends on the lepton flavors in the final state.
In this study, we consider two scenarios: different flavor charged
lepton pairs, i.e. $H\to2e2\mu$, and four same flavor charged leptons,
i.e. $H\to4e(\,{\rm or}\,\,4\mu)$. Hence, we have two methods for
reconstructing the $Z$ bosons:

\begin{enumerate}
\item Different flavor charged lepton pairs ($2e2\mu$):\\
 In this case, it is easy to reconstruct the $Z$ bosons because both
electron and muon lepton flavors can be tagged. Using the flavor information,
the $Z$ bosons can be reconstructed by summing over the same flavor
opposite-sign leptons in the final state. 
\item Four same flavor charged leptons ($4e/4\mu$): \\
 If the flavors of four leptons are all the same, one needs to pursue
some algorithms to reconstruct the $Z$ boson mass. In our analysis,
we first pair up the leptons with opposite charge. We require the
pair whose invariant mass is closest to $M_{Z}$ to be the one generated
from the on-shell $Z$ boson, and the other pair is the one generated
from another $Z$ boson, which could be on-sell or off-shell. We name
it as the minimal deviation algorithm (MDA) in this paper. 
\end{enumerate}
In Fig.~\ref{fig:zrec}, we show the $p_{T}$ distributions of the
reconstructed $Z$ boson for $140$ and $200\,{\rm GeV}$, respectively.
When the final state lepton flavors are different, one can reconstructed
the $Z$ boson perfectly by matching the lepton flavor. For the same
flavor leptons, the reconstructed $Z$ boson distributions in the
MDA are shown as the solid, dashed and dot-dashed curves for RES,
NLO and LO, respectively. Some points are worthy to point out as follow:

\begin{itemize}
\item We note that the MDA can perfectly reconstruct the distributions of
true $Z$ bosons, irregardless whether these two $Z$ bosons are both
on-shell or only one of them is on-shell. 
\item When $M_{H}=200\,{\rm GeV}$, both $Z$ bosons are produced on-shell
and boosted. The peak position of the transverse of momentum $p_{T}^{Z}$
is around $\sqrt{\left(M_{H}/2\right)^{2}-m_{Z}^{2}}\sim41\,{\rm GeV}$.
For all the cases, the RES effects change the shape of $p_{T}^{Z}$
largely and shift the $p_{T}^{Z}$ to the larger value region. 
\end{itemize}

\begin{figure}
\includegraphics[clip,scale=0.6]{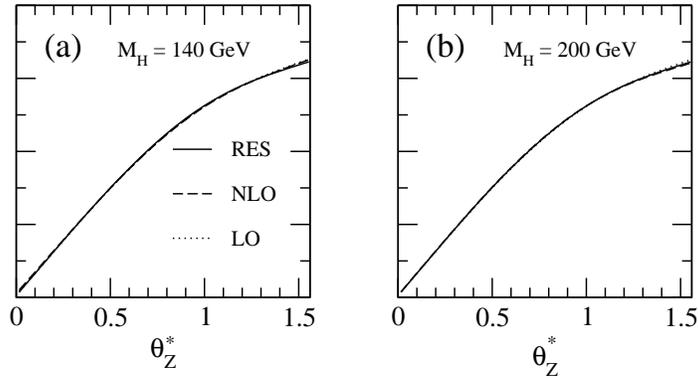}

\caption{Normalized polar angle of the (back-to-back) $Z$ boson momenta distributions
in the rest frame of the Higgs boson in $gg\to H\to ZZ^{(*)}\to\ell^{+}\ell^{-}\ell^{\prime+}\ell^{\prime-}$
at the LHC: (a) is for $M_{H}=140\,{\rm GeV}$ , (b) is for $M_{H}=200\,{\rm GeV}$.
\label{fig:Z-star}}
\end{figure}

\begin{figure}
\includegraphics[clip,scale=0.6]{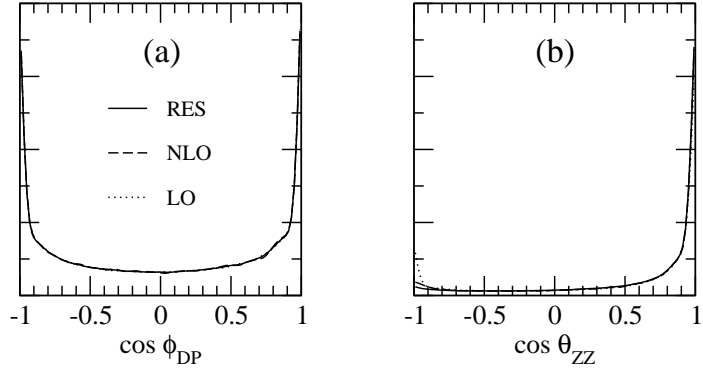}

\caption{Normalized distributions of $\cos\phi_{DP}$ and $\cos\theta_{ZZ}$
in the rest frame of the Higgs boson with mass $200\,{\rm GeV}$ in
$gg\to H\to ZZ\to\ell^{+}\ell^{-}\ell^{\prime+}\ell^{\prime-}$ at
the LHC.\label{fig:cth-phiplane}}
\end{figure}

It has been shown in Ref.~\cite{Mellado:2004} that angular correlation
between the two $Z$ bosons from the Higgs decay can be used to suppress
the intrinsic background from $ZZ$ pair production efficiently. One
of the useful angular variables is the polar angle ($\theta_{Z}^{*}$)
of the (back-to-back) $Z$ boson momenta in the rest frame of the
Higgs boson~\cite{Mellado:2004}. As shown in Fig.~\ref{fig:Z-star},
in the rest frame of Higgs boson, the back-to-back $Z$ bosons like
to lie in the direction perpendicular to the $z-$axis, which is the
moving direction of the Higgs boson in the lab frame. After being
boosted to the lab frame, two $Z$ bosons will move close to each
other, c.f. Fig.~\ref{fig:cth-phiplane}(b), where $\theta_{ZZ}$
is the opening angle between the two $Z$ bosons in the lab frame.
Another interesting angular variable is the angle between the two
on-shell $Z$ boson decay planes ($\phi_{DP}$) in the rest frame
of the Higgs boson, which is shown in Fig.~\ref{fig:cth-phiplane}(a).
The two $Z$ bosons are reconstructed as explained above. Since the
angle $\theta_{Z}^{*}$ and $\phi_{DP}$ are defined in the rest frame
of the Higgs boson, the non-zero transverse momentum of the Higgs
boson does not affect these two variables. Therefore, as clearly shown
in the figures, all the distributions of the angular variables mentioned
above are the same for the RES, NLO and LO calculations.

\textcolor{black}{}%

\subsection{$gg\to H\to ZZ\to\ell^{+}\ell^{-}\nu\bar{\nu}$}

\begin{figure}
\includegraphics[clip,scale=0.7]{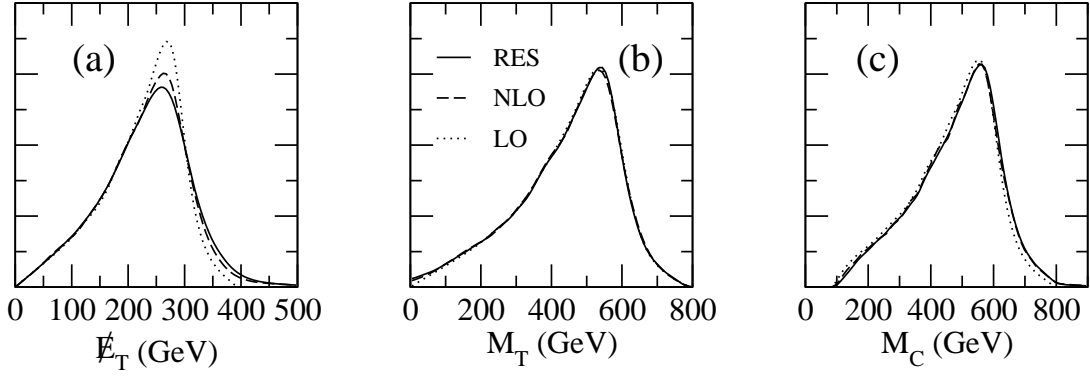}

\caption{Normalized distributions of $\met$, $M_{T}$ and $M_{C}$ in $gg\to H\to ZZ\to\ell^{+}\ell^{-}\nu\nu$
channel with $M_{H}=600\,{\rm GeV}$ at the LHC.~\label{fig:MisET_zz-MTC}}
\end{figure}

Although the {}``gold-plated'' mode, $H\to ZZ\to\ell^{+}\ell^{-}\ell^{\prime+}\ell^{\prime-}$,
is considered to be the most effective channel for the SM Higgs boson
discovery at the LHC, it suffers from the small decay branching of
$Z\to\ell^{+}\ell^{-}$. Moreover, the larger the Higgs mass becomes,
the smaller the production rate is. When the Higgs boson mass is larger
than 600\ GeV, the $H\to ZZ\to\ell^{+}\ell^{-}\nu\bar{\nu}$ channel
may become important because the decay branching ratio (Br) of $H\to ZZ\to\ell^{+}\ell^{-}\nu\bar{\nu}$
is six times of the Br of $H\to ZZ\to\ell^{+}\ell^{-}\ell^{\prime+}\ell^{\prime-}$.
The drawback is that one cannot reconstruct the Higgs mass from the
final state particles due to the presence of two neutrinos. In this
discovery channel, the missing transverse energy ($\met$) is crucial
to suppress the background~\cite{atlas:1999}. The $\met$ distribution
is shown in Fig.~\ref{fig:MisET_zz-MTC}(a) which exhibits a Jacobian
peak around $M_{H}/2$, and the soft-gluon resummation effects smear
the Jacobian peak and shift more events to the larger $\met$ region.
Similar to the $H\to WW$ mode, the kinematics of this channel is
similar to the $W$ boson production and decay in the Drell-Yan process,
therefore the shape of $\met$ distribution change significantly by
the RES contributions. The Higgs boson mass can be measured from the
peaks of the distributions of the transverse mass $M_{T}$ and the
cluster mass $M_{C}$, cf. Eq.~(\ref{eq:mt-mc}), as shown in Fig.~\ref{fig:MisET_zz-MTC}(b)\ and\ (c).
Although the upper endpoint of $M_{T}$ is insensitive to high order
corrections as we mentioned in the study of $H\to WW^{(*)}$ mode,
the Jacobian peak is smeared out by the width ($\Gamma_{H}$) effects
of the Higgs boson. For $M_{H}=600\,{\rm GeV}$, the total decay width
of the Higgs boson is about 120 GeV, which is quite sizable and generates
a noticeable smearing effect on the Jacobian peak.

\subsection{Acceptance study}

\begin{table}

\caption{Acceptance of the process $gg\to H\to ZZ\to\ell^{+}\ell^{-}\ell^{\prime+}\ell^{\prime-}$
for $M_{H}=140\,(200)\,{\rm GeV}$ and the process $gg\to H\to ZZ\to\ell^{+}\ell^{-}\nu\bar{\nu}$
for $M_{H}=600\,{\rm GeV}$ after imposing cuts.\label{tab:z-Accept}}

\begin{tabular}{c|c|c|c|c|c}
\hline 
&
\multicolumn{2}{c|}{$M_{H}=140\,{\rm GeV}$}&
\multicolumn{2}{c|}{$M_{H}=200\,{\rm GeV}$}&
\multicolumn{1}{c}{$M_{H}=600\,{\rm GeV}$}\tabularnewline
\hline
\hline 
&
basic (Eq.\ \ref{eq:z140cut1})&
optimal (Eq.\ \ref{eq:z140cut2})&
basic (Eq.\ \ref{eq:z140cut1})&
optimal (Eq.\ \ref{eq:z140cut2})&
basic (Eq.\ \ref{eq:z600cut})\tabularnewline
\hline 
RES&
0.53&
0.15&
0.67&
0.14&
0.55\tabularnewline
NLO&
0.54&
0.12&
0.67&
0.11&
0.56\tabularnewline
LO&
0.53&
0&
0.67&
0&
0.58\tabularnewline
\hline
\end{tabular}
\end{table}

The discovery potential of the $H\to ZZ\to\ell^{+}\ell^{-}\ell^{\prime+}\ell^{\prime-}$
and $H\to ZZ\to\ell^{+}\ell^{-}\nu\bar{\nu}$ modes has been studied
in Ref.~\cite{atlas:1999} after imposing the following cuts:

\begin{itemize}
\item For $M_{H}=140\,{\rm GeV}$ and $200\,{\rm GeV}$, the intermediate
mass range, we impose the \emph{basic cuts}:\begin{eqnarray}
p_{T}^{E}>7.0\,{\rm GeV}, & \left|Y_{L}\right|<2.5, & p_{T}^{L}>20\,{\rm GeV},\label{eq:z140cut1}\end{eqnarray}
 and the \emph{optimal cuts}:\begin{equation}
p_{T}^{Z_{Max}}>\frac{M_{H}}{3},\label{eq:z140cut2}\end{equation}
 where $p_{T}^{E}$ and $Y_{L}$ are the transverse momentum and rapidity
of each charged lepton, respectively, and $p_{T}^{Z_{Max}}$ is the
$p_{T}$ of the harder $Z$ boson. 
\item For $M_{H}=600\,{\rm GeV}$, we require:\begin{eqnarray}
p_{T}^{L}>40\,{\rm GeV}, &  & \left|Y_{L}\right|<2.5\,,\nonumber \\
p_{T}^{LL}>200\,{\rm GeV}, &  & \met>150\,{\rm GeV},\label{eq:z600cut}\end{eqnarray}

\end{itemize}
where $p_{T}^{LL}$ is the transverse momentum of the two charged
lepton system. The numerical results of the acceptances of the various
cuts are summarized in Table\ \ref{tab:z-Accept}. For the $H\to ZZ^{(*)}\to\ell^{+}\ell^{-}\ell^{\prime+}\ell^{\prime-}$
mode, the RES and NLO contributions have almost the same acceptances
after imposing the basic cuts. However, after imposing the optimal
cuts the acceptance of the RES contribution is larger than the one
of the NLO contribution by $25\%$, and the LO contribution is largely
suppressed. For the $H\to ZZ\to\ell^{+}\ell^{-}\nu\bar{\nu}$ mode,
the acceptances of the RES and NLO calculations are similar to each
other.

\section{Conclusion\label{sec:Conclusion}}

The search for the SM Higgs boson is one of the major goals of the
high energy physics experiments at the LHC, and the vector boson decay
modes, $H\to WW^{(*)}$ or $H\to ZZ^{(*)}$, provide powerful and
reliable discovery channels. The LHC has a great potential to discover
the Higgs boson even with low luminosity ($\sim30\,{\rm fb}^{-1}$)
during the early years of running~\cite{atlas:1999,CMS:2006,Pieri:2005ud}.
In order to extract the signal from huge background events, we should
have better theoretical predictions of the signal events as well as
background events. In this paper, we examine the soft gluon resummation
effects on the search of SM Higgs boson via the dominant production
process $gg\to H$ at the LHC and discuss the impacts of the resummation
effects on various kinematics variables which are relevant to the
Higgs search. A comparison between the resummation effects and the
NLO calculation is also presented.

For $H\to WW^{(*)}\to\ell^{+}\ell^{-}\nu\bar{\nu}$ mode, we study
$M_{H}=140\,{\rm GeV}$ at the Tevatron and $M_{H}=170\,{\rm GeV}$
at the LHC. Due to the spin correlations between the final state particles,
this process is similar to the $W$ boson production and decay in
the Drell-Yan process. The shapes of the kinematics distributions
are modified significantly by RES effects. For example, the effects
could cause $\sim50\%$ difference compared to NLO calculation \textcolor{black}{in
the transverse momentum distribution of the leading lepton} ($p_{T}^{L_{Max}}$),
when $M_{H}=170\,{\rm GeV}$. The Higgs boson mass cannot be reconstructed
directly from the final state particles because of two neutrinos.
Therefore, the upper endpoint in the transverse mass distribution
can be used to determine the mass of the Higgs boson, and we found
that it is insensitive to the RES effects. After imposing various
kinematics cuts, the LO, NLO and RES calculations \textcolor{black}{yield
similar acceptance of the signal events}.

For the $H\to ZZ^{(*)}\to\ell^{+}\ell^{-}\ell^{\prime+}\ell^{\prime-}$
mode, the so-called {}``gold-plated'' mode, we study $M_{H}=140\,{\rm GeV}$
and $M_{H}=200\,{\rm GeV}$ at the LHC in this paper. We pursue an
algorithm, called minimal deviation algorithm in this paper, to reconstruct
the two $Z$ bosons when the four charged leptons in the final state
have the same flavors. The RES effects change the shapes of kinematics
significantly, e.g. $p_{T}^{L_{max}}$ and $p_{T}^{Z}$ distributions.
However, the variables $\phi_{DP}$ and $\theta_{Z}^{*}$, defined
in the Higgs rest frame, are insensitive to RES effects. After imposing
the optimal kinematics cuts, the RES effects could increase the acceptance
by $25\%$ compared to that of NLO calculation while \textcolor{black}{the
LO contribution is largely suppressed}. When the Higgs boson is heavy
($600\,{\rm GeV}$), we consider the $H\to ZZ\to\ell^{+}\ell^{-}\nu\bar{\nu}$
mode because of its larger decay branching ratio, as compared to the
$H\to ZZ\to\ell^{+}\ell^{-}\ell^{\prime+}\ell^{\prime-}$ mode. The
shape of $\met$ distribution, which is crucial to suppress the backgrounds,
is largely modified because it is sensitive to the transverse momentum
of the Higgs boson.

\textcolor{black}{In summary, we have presented a study of initial
state soft-gluon resummation effects on the search for the SM Higgs
boson via gluon-gluon fusion at the LHC. The effects not only significantly
modify some of the kinematic distributions of the final state particles,
as compared to the NLO and LO predictions, but also enhance the acceptance
of the signal events after imposing the kinematic cuts to suppress
the large background events. Therefore, we conclude that the initial
state soft-gluon resummation effects should be taken into account
as searching for the Higgs boson at the LHC. In addition, we note
that the spin correlations among the final state leptons could be
modified by the electroweak corrections to the Higgs boson decay.
Therefore, we have implemented the NLO QED correction in the ResBos
code, and the phenomenological study will be presented in the forthcoming
paper.} \textcolor{red}{}

\begin{acknowledgments}
We thank Professor C.-P. Yuan for a critical reading and useful suggestions.
We also thank Dr. Kazuhiro Tobe for useful discussions. Q.-H. Cao
is supported in part by the U.S. Department of Energy under grant
No. DE-FG03-94ER40837. C.-R. Chen is supported in part by the U.S.
National Science Foundation under award PHY-0555545. \bibliographystyle{apsrev}
\bibliographystyle{apsrev}
\bibliography{hvv}

\end{acknowledgments}

\end{document}